\documentclass[letterpaper,aps,prd,twocolumn,floats,showpacs,preprintnumbers,nofootinbib,superscriptaddress]{revtex4}

\usepackage{amsmath,amssymb}
\usepackage{epsfig}
\usepackage{bbm}

\newcommand{\D}{{\rm d}}

\begin{document}
\preprint{YITP-SB-09-35}
\title{Quantum Decoherence of Photons in the Presence of Hidden U(1)s}

\author{M.~Ahlers}

\affiliation{C.~N.~Yang Institute for Theoretical Physics, 
State University of New York at Stony Brook, 
NY 11794-3840, USA}

\author{L.~A.~Anchordoqui}

\affiliation{Department of Physics, University of Wisconsin-Milwaukee, 
Milwaukee, WI 53201, USA}

\author{M.~C.~Gonzalez--Garcia}

\affiliation{Instituci\'o Catalana de Recerca i Estudis Avan\c{c}ats (ICREA), 
Departament d'Estructura i Constituents de la Mat\`eria, 
Universitat de Barcelona, 
647 Diagonal, E-08028 Barcelona, Spain}
\affiliation{C.~N.~Yang Institute for Theoretical Physics, 
State University of New York at Stony Brook, 
NY 11794-3840, USA}

\begin{abstract}
Many extensions of the standard model predict the existence of hidden
sectors that may contain unbroken abelian gauge groups. We argue that
in the presence of quantum decoherence photons may convert into hidden
photons on sufficiently long time scales and show that this effect is
strongly constrained by CMB and supernova data. In particular, Planck-scale suppressed decoherence scales
 $D \propto \omega^2/M_{\rm Pl}$ (characteristic for non-critical string theories) are incompatible with the presence of even a single hidden U$(1)$. The corresponding bounds on the decoherence scale are four orders of magnitude stronger than analogous bounds derived from solar and reactor neutrino data and complement other bounds derived from atmospheric neutrino data.
\end{abstract} 

\pacs{14.80.-j, 95.36.+x, 98.80.Es}

\maketitle

\section{Motivation}

Physics beyond the standard model typically predicts the existence of hidden
sectors containing extra matter with new gauge interactions. There is no
compelling reason why these extra sectors should be very massive if their
interaction with standard model matter is sufficiently weak. This can be
accomplished, {\it e.g.}, in extra dimensional extensions with a geometric
separation of sectors or in models where tree-level interactions are absent and
higher order corrections suppressed by the scale of messenger masses.

The presence of these light hidden sectors can have various observable
effects.  For example, in the case of kinetic
mixing~\cite{Holdom:1985ag,Dienes:1996zr} between a hidden U$(1)$ and
the electromagnetic U$(1)$, hidden sector matter can receive small
fractional electro-magnetic charges. These {\it mini-charged}
particles can have a strong influence on early Universe physics and
astrophysical environments~\cite{Raffelt:1996wa}. If the hidden sector
U$(1)$ is slightly broken by a Higgs or St\"uckelberg mechanism we can
have a situation analogous to neutrino systems with characteristic
oscillation patterns between photons and hidden photons over
sufficiently long baselines~\cite{Okun:1982xi}. 

Here we concentrate on another effect induced by the presence of
these light hidden sectors: the sensitivity of photon propagation to
sources of quantum decoherence, {\it e.g.}~quantum gravity
effects~\cite{Hawking:1982dj}.  
A heuristic picture describes space-time at the Planck
scale as a foamy structure~\cite{Wheeler:1957mu}, where virtual black holes pop in and out
of existence on a time scale allowed by Heisenberg's uncertainty
principle~\cite{Hawking:1995ag}. This can lead to a loss of quantum
information across their event horizons, providing an ``environment''
that might induce quantum decoherence of apparently isolated matter
systems~\cite{Hawking:1982dj}.

It is an open matter of debate whether quantum decoherence induced by
a quantum theory of gravity would simultaneously preserve Poincare
invariance and
locality~\cite{Hawking:1982dj,Hawking:1995ag,Ellis:1983jz,Banks:1983by,Unruh:1995gn}.
A violation of energy and momentum conservation by particle reactions
with a space-time foam could be reflected by an (energy dependent)
effective refractive index in vacuo~\cite{AmelinoCamelia:1996pj}. This
could be tested, {\it e.g.}, by the measurement of the arrival time of
gamma rays or high energy neutrinos at different energies or by the
propagation of ultra-high energy cosmic rays \cite{GonzalezMestres:1997dr}.

If, on the other hand, Poincare invariance is preserved, the presence of
a non-trivial space-time vacuum can still be signaled by decoherence
effects in systems of stable elementary particles  \cite{Ellis:1983jz}. 
A particularly interesting and well-studied case are neutrino systems, 
where the interplay between mixing, mass oscillation and decoherence can
influence atmospheric, solar and reactor neutrino data 
\cite{Lisi:2000zt,Gago:2002na,Fogli:2007tx}, as well
as flavour composition of astrophysical 
high-energy neutrino fluxes~\cite{Hooper:2004xr, Anchordoqui:2005gj}.

If hidden sectors contain unbroken abelian gauge groups it is also
feasible that the system of the electromagnetic photon ($\gamma_0$)
and hidden photons ($\lbrace\gamma_i\rbrace$ with $i\geq1$) experience
energy and momentum conserving decoherence effects:  transitions between different photon ``species'',
$\gamma_\mu\to\gamma_\nu$, are allowed by gauge and Poincare
invariance. This is the only alternative system to neutrinos
involving a stable and neutral elementary particle of the standard
model, which can experience these decoherence effects on extremely long
(cosmological) time scales.

The outline of this paper is as follows. We will start in
Sect.~\ref{sec:QD} with an outline of the Lindblad formalism 
of quantum decoherence. We will
then discuss in Sect.~\ref{sec:absorption} the effect of photon
decoherence on the Planck spectrum of the cosmic microwave background
and the luminosity distance of type Ia supernovae, respectively. This
enables us to derive strong limits on various decoherence models. We
comment in Sect.~\ref{sec:propagation} on the interplay of decoherence
and photon interactions and outline a possible mechanism to extend the
survival probability of extra-galactic TeV gamma rays. We finally
summarize in Sect.~\ref{sec:summary}.

\section{Quantum Decoherence}\label{sec:QD}

The Lindblad formalism is a general approach to quantum decoherence
that does not require any detailed knowledge of the
environment~\cite{Lindblad:1975ef}. In the presence of decoherence the
modified Liouville equation can be written in the form
\begin{equation}\label{eq:liouville_mod}
\frac{\partial \rho}{\partial t} = -{\rm i}[H,\rho] +
\mathcal{D}[\rho]\,.
\end{equation}
The Hamiltonian $H$ can include possible background contributions,
{\it e.g.}~plasma effects for the photon. However, this does not
effect the evolution of the density matrix in the absence of mixing
between the gauge bosons. The decoherence term $\mathcal{D}$ in the
modified Liouville equation~(\ref{eq:liouville_mod}) can be written as
\begin{equation}\label{eq:lindblad}
{\cal D} [\rho] = \frac{1}{2} \sum_j \left([b_j,\, \rho\, b_j^\dagger]
+ [b_j\, \rho,\, b_j^\dagger]\right)\,,
\end{equation}
where $\lbrace b_j\rbrace$ is a sequence of bounded operators acting
on the Hilbert space of the open quantum system, ${\cal H}$, and
satisfying $\sum_j b^\dagger_j b_j \in {\cal B} ({\cal H}),$ where
${\cal B} ({\cal H})$ indicates the space of bounded operators acting
on ${\cal H}$. The dynamical effects of spacetime on a microscopic
system can then be interpreted as the existence of an arrow of time
which in turn makes possible the connection with thermodynamics via an
entropy. The monotonic increase of the von Neumann entropy, $S(\rho) =
- {\rm Tr}(\rho \ln \rho)$, implies the hermiticity of the Lindblad
operators, $b_j = b_j^\dagger$~\cite{Benatti:1987dz}. In addition, the
conservation of energy and momentum can be enforced by taking $[P_\mu,
b_j] = 0$.

We will assume in the following that there is a total of $N$ abelian
gauge bosons $\gamma_\mu$ with $\mu=0,\ldots,N-1$ including the photon
$\gamma_0$. The solution to Eq.~(\ref{eq:liouville_mod}) is outlined
in Appendix~\ref{app:lindblad}.  For simplicity, we assume degeneracy
of the decoherence parameters $D_i=D$ which simplifies the photon
survival probability after a distance $x=t=L$ significantly (see
App.~\ref{app:lindblad}),
\begin{equation}\label{eq:Pgg}
P_{\gamma\to\gamma} = \frac{1}{N}+\frac{N-1}{N}e^{-DL}\,.
\end{equation}

The energy behavior of $D$ depends on the dimensionality of the
operators $b_j.$ We can estimate the energy dependence from gauge
invariance. Possible combinations of the field strength tensors $F_{\mu\nu}$
and $G_{\mu\nu}$ of the two U$(1)$s are~\cite{Hawking:1982dj} $b_j \propto(F_{\mu\nu}G^{\mu\nu})^j \propto\omega^j$.
This restriction of the energy behavior to non-negative powers of
$\omega$ may possibly be relaxed when the dissipative term is directly
calculated in the most general space-time foam
background~\cite{Anchordoqui:2005gj}.

An interesting example is the case where the dissipative term is
dominated by the dimension-4 operator $b_1$ yielding the energy
dependence $D \propto \omega^{2}/M_{\rm Pl}.$ This is characteristic
of non-critical string theories where the space-time defects of the
quantum gravitational ``environment'' are taken as recoiling
$D$-branes, which generate a cellular structure in the space-time
manifold~\cite{Ellis:1996bz}.

\section{Observational Constraints}\label{sec:absorption}

In the following we will investigate the limits on quantum decoherence
in the presence of hidden {\it massless} U$(1)$s from cosmological and
astrophysical observations. Unless otherwise stated, we will make the
conservative assumption that there exists only a single hidden U$(1)$
in addition to the standard model ($N=2$) and that the decoherence
rate of photons can be parametrized as
\begin{align}
\label{eq:GIG}
D(z,\omega) = (1+z)^{p+1}D_*\frac{\omega^{p+1}}{M^{p}_{\rm QD}}\,,
\end{align}
where $M_{\rm QD}$ is the scale of quantum decoherence, not
necessarily the Planck scale. We will consider the scale dependencies
$p=0,\pm1,\pm2$.

The differential flux of photons from a source at redshift $z$ is
reduced by the exponential factor
\begin{equation}
P_{\gamma\to\gamma}(z,\omega) =
\frac{1}{2}+\frac{1}{2}\exp\left[-\int_0^z\D\ell'\,D(z',\omega)\right]\,,
\end{equation}
where the propagation distance $\ell$ is given by \mbox{$\D z =
H(z)(1+z)\D \ell$} with Hubble parameter $H$. The Hubble parameter at
redshift $z$ is given by \mbox{$H^2(z) =
H_0^2[(1-\Omega_m-\Omega_r)+\Omega_m(1+z)^3+\Omega_r(1+z)^4]$} where
$\Omega_mh^2 \simeq 0.128$ and $\Omega_rh^2 \simeq
2.47\times10^{-5}$. The present Hubble expansion is \mbox{$H_0 =
h\,100\,{\rm km}\,{\rm s}^{-1}\,{\rm Mpc}^{-1}$} with
$h\simeq0.73$~\cite{Amsler:2008zzb}.

%%%%%%%%%%%%%%%%%%%%%%%%%%%%%%
\begin{figure}[t!]
\begin{center}
\hspace{0.15cm}\includegraphics[width =0.98\linewidth]{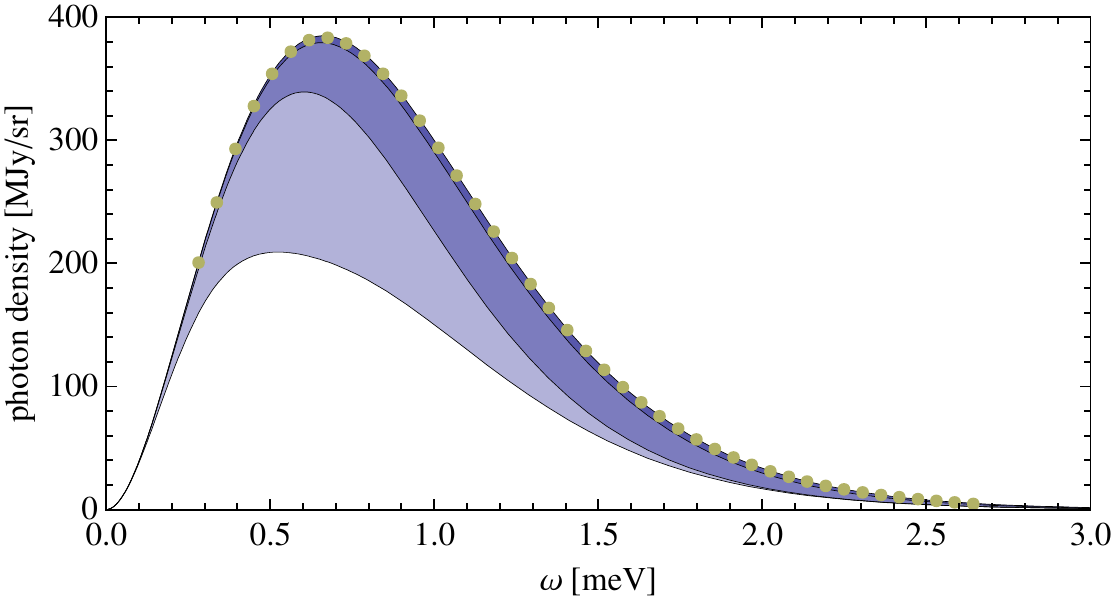}\\[0.2cm]
\includegraphics[width =\linewidth]{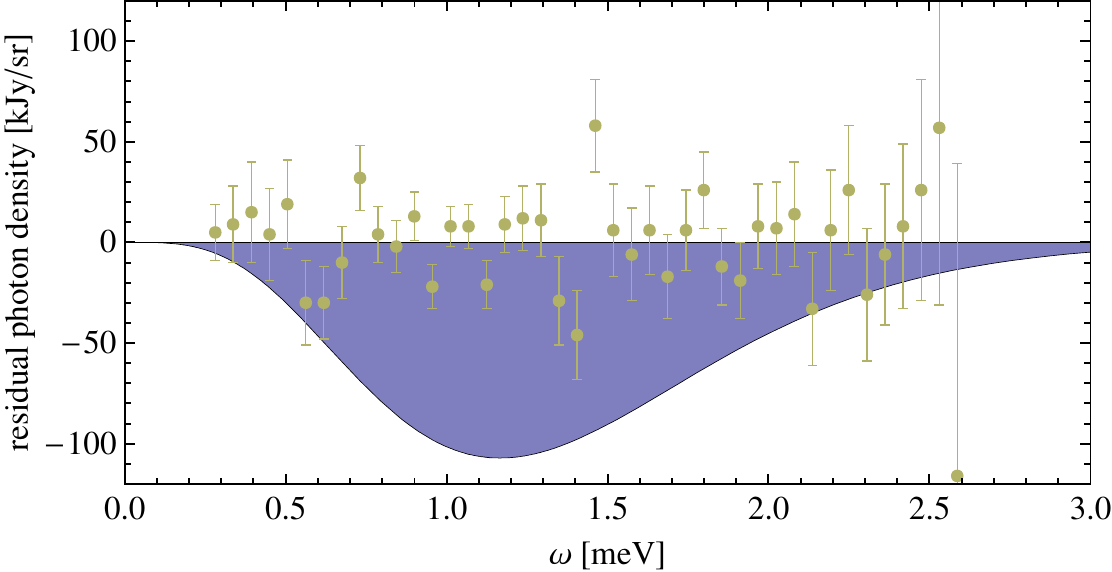}
\end{center}
\vspace{-0.3cm}
\caption[]{Modification of the CMB spectrum by decoherence
effects. The dots show the residual CMB spectrum measured by
COBE/FIRAS~\cite{FIRAS} ($1$Jy (Janksy)
$\equiv10^{-26}$~W~m${}^{-1}$~Hz${}^{-1}$). The upper panel shows the
effect of $D_*=0.01,0.1,1$ and the lower panel $D_*=10^{-4}$. The
curves show the accumulated deficit in photons compared to the ``no
decoherence'' case. (A different shadowing is included to guide the eye.)}
\label{fig:planck}
\end{figure}
%%%%%%%%%%%%%%%%%%%%%%%%%%%%%%
%%%%%%%%%%%%%%%%%%%%%%%%%%%%%%
\begin{table*}[ph!]
\begin{minipage}{0.76\linewidth}
\begin{center}
\setlength{\tabcolsep}{8pt}\renewcommand{\arraystretch}{1.4}
\begin{tabular}{c|cc|cc}\hline\hline
Model&CMB&SNe&reactor\&solar $\nu$&atmospheric $\nu$\\
\hline
\,\,$p=-2$\,\,&\,\,\,\,$9.5\times10^{-97}$
\,\,\,\,&\,\,\,\,$3.9\times10^{-90}$
\,\,\,\,&$8.3\times10^{-65}$~\footnote{\protect \cite{Fogli:2007tx} $3\sigma$ C.L.~from their 
Fig.~1 with $n=p+1$ and $\gamma_0/{\rm GeV} = D_* (M_{\rm QD}/{\rm GeV})^p$.}
&$1.0\times10^{-59}$~\footnote{\cite{Lisi:2000zt} The authors consider the case $p=-2$ with 
$\mu^2 = 2D_*M_{\rm QD}^2$}
\\
$p=-1$&$9.7\times10^{-66}$&$3.4\times10^{-62}$&$8.8\times10^{-44}$~\footnotemark[1]
&-\\
$p=0$&$3.9\times10^{-35}$&$2.9\times10^{-34}$&
$1.0\times10^{-22}$~\footnotemark[1]
&$1.2\times10^{-27}$~\footnote{\cite{Collaboration:2009nf} Bounds are 
at the $90$\%~C.L. Their notation corresponds to
$n=p+1$ and $d_* = D_* (M_{\rm pl}/M_{\rm QD})^p$.}   
\\
$p=1$&$1.3\times10^{-5}$&$2.3\times10^{-6}$&
$0.11$~\footnotemark[1]
&$1.6\times10^{-12}$~\footnotemark[3]\\
$p=2$&$3.5\times10^{23}$&$1.8\times10^{22}$&-&$910$~\footnotemark[3]\\\hline\hline
\end{tabular}
\end{center}
\end{minipage}
\caption[]{The $3\sigma$ limits on $D_*$ derived from the COBE/FIRAS
data~\cite{FIRAS} and SN data~\cite{Kowalski:2008ez} for a decoherence scale
given by Eq.~(\ref{eq:GIG}). We assume $N=2$ and $M_{\rm QD}=M_{\rm Pl}$. Note
that the limits are in general stronger for $N>2$ by up to a factor $2$. 
For comparison the last two columns show the limits from reactor and 
solar ~\cite{Fogli:2007tx} , and atmospheric neutrinos ~\cite{Lisi:2000zt,Collaboration:2009nf}.}
\label{tab:threesigma}
\end{table*}
%%%%%%%%%%%%%%%%%%%%%%%%%%%%%%

\subsection{CMB Distortions}\label{sec:CMB}

In the standard big bang cosmology the cosmic microwave background
(CMB) forms at a redshift of about $z_{\rm CMB}\simeq1100$ after
recombination of electrons and (mostly) protons in the expanding and
cooling Universe. The CMB is well described by a Planck spectrum with
a temperature of $T= 2.725 \pm 0.001$~K~\cite{FIRAS},
\begin{equation}
\frac{\D^2
n}{\D\omega\D\Omega}=\frac{1}{2\pi^2}
\frac{\omega^3}{\exp\left(\frac{\omega}{T}\right)-1}\,.
\end{equation}
The high degree of accuracy (better than $1$ in $10^4$ around $1$~meV)
between the CMB measurement and cosmological predictions is an ideal
probe for exotic physics that could have effected the CMB photons in
the redshift range $0<z<z_{\rm CMB}$ with energies from meV ($z=0$) to
eV ($z=z_{\rm CMB}$).

The influence of light particles coupling to the CMB has been studied
previously for the case of axion-like particles~\cite{Mirizzi:2005ng},
mini-charged particles~\cite{Melchiorri:2007sq} and massive hidden
photons with kinetic mixing~\cite{Jaeckel:2008fi}. In the case of
photon absorption or decoherence the observed spectrum is modified as
\begin{equation}
\frac{\D^2 n^{\rm obs}}{\D\omega\D\Omega}=P_{\gamma\to\gamma}(z_{\rm
CMB},\omega)\,\frac{\D^2 n}{\D\omega\D\Omega}\,.
\end{equation}

As an illustration of the effect of decoherence, Fig.~\ref{fig:planck}
shows the distortion of the CMB spectrum for the case $p=1$ and
$M_{\rm QD}=M_{\rm Pl}\simeq1.2\times10^{19}$~GeV for $D_*=0.01,0.1,1$
(upper panel) and $D_*=10^{-4}$ (lower panel). The $3\sigma$ limits
from a $\chi^2$-fit of various models to the COBE/FIRAS
data~\cite{FIRAS} are shown in Tab.~\ref{tab:threesigma} (see also 
upper left panel in Fig.~\ref{fig:exclusion}). For
comparison, the two right columns show limits on the decoherence
parameter derived from solar and reactor neutrino
data~\cite{Fogli:2007tx} and atmospheric neutrino
data~\cite{Lisi:2000zt,Collaboration:2009nf}. We will comment on how
to compare these bounds at the end of the section. 

\subsection{SN Dimming}\label{sec:SNdimming}

Also the luminosity distance measurements of cosmological standard
candles like type Ia supernovae
(SNe)~\cite{Riess:1998cb,Kowalski:2008ez} are able to test feeble photon
absorption and decoherence effects. The luminosity distance $d_L$ is
defined as
\begin{equation}
\label{dL}
d_L(z) \equiv \sqrt{\frac{\mathcal{L}}{4\pi F}}\,,
\end{equation}
where $\mathcal{L}$ is the luminosity of the standard candle (assumed
to be sufficiently well-known) and $F$ the measured flux. In a
homogeneous and isotropic universe this is predicted to be
\begin{equation}
d_L(z) = (1+z)\,a_0\,\Phi
\left( \int_0^z\frac{\D z'}{a_0H(z')}\right)\,,
\end{equation}
with $a_0^{-1} = H_0\sqrt{|1-\Omega_{\rm tot}|}$ and
$\Phi_k(\xi)=(\sinh\xi,\xi,\sin\xi)$ for spatial curvature
\mbox{$k=-1,0,1$}, respectively. If the photon flux of a source,
located at distance $z$ and observed in a (small) frequency band
centered at $\omega_\star$, is attenuated by photon interactions or
quantum decoherence the observed luminosity distance {\it increases}
as
\begin{equation}
d_L^{\,\rm obs}(z) =
\frac{d_L(z)}{\sqrt{P_{\gamma\to\gamma}(z,\omega_\star)}}\,.
\end{equation}

The apparent extension of the luminosity distance by photon
interactions and oscillations has been investigated in the context of
axion-like
particles~\cite{Csaki:2001yk,Mirizzi:2005ng}, hidden
photons~\cite{Evslin:2005hi}, chameleons~\cite{Burrage:2007ew} and
mini-charged particles~\cite{Ahlers:2009kh}. One of the main
attractions of these models is the possibility that the conclusions
about the energy content of our Universe drawn from the Hubble diagram
can be dramatically altered. We will briefly comment on this
possibility at the end of this section.

%%%%%%%%%%%%%%%%%%%%%%%%%%%%%%
\begin{figure}[t!]
\begin{center}
\includegraphics[width = \linewidth]{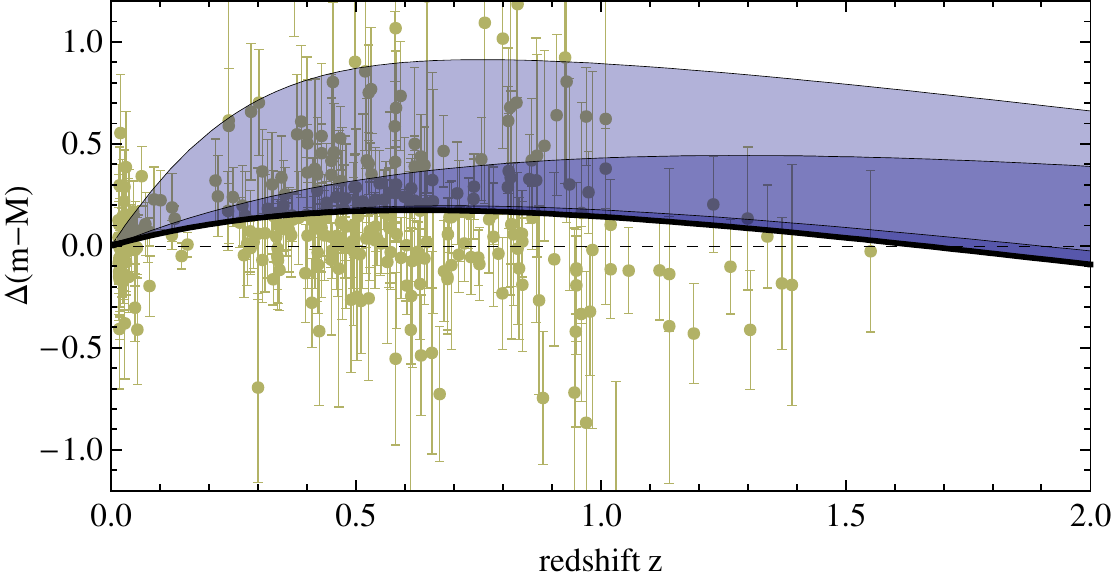}\\[0.2cm]
\includegraphics[width = \linewidth]{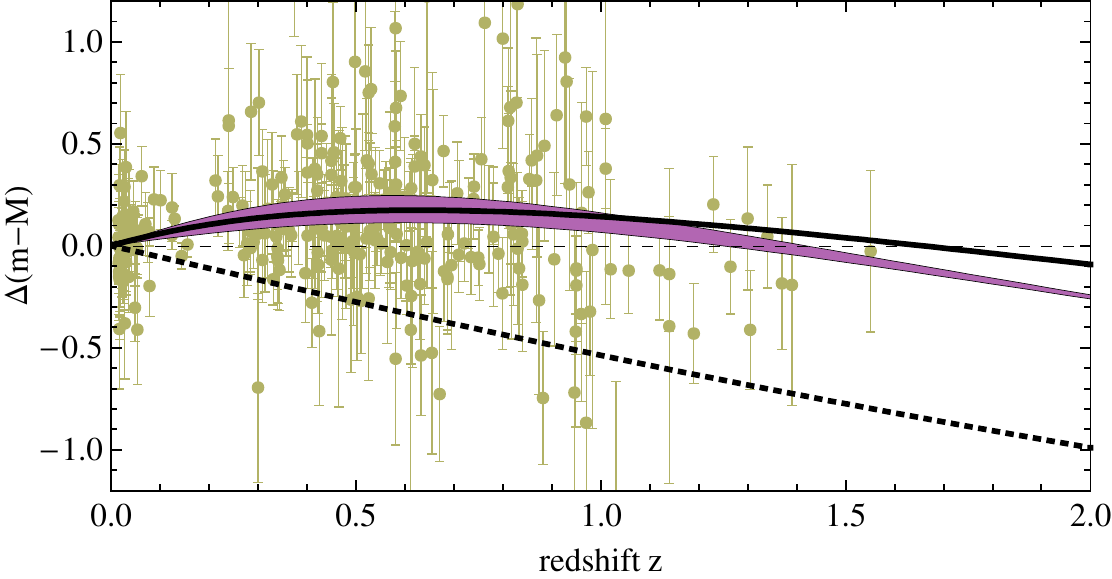}
\end{center}
\vspace{-0.3cm}
\caption[]{{\bf Upper Panel:} Enhanced SNe dimming by decoherence
effects assuming $N=2$, $p=1$, $M_{\rm QD}=M_{\rm Pl}$ and
$D_*=10^{-6},10^{-5},10^{-4}$ assuming sources observed in a frequency
interval centered at $\omega_\star=1$~eV. The dots show the SNe Ia
``union'' compilation from Ref.~\cite{Kowalski:2008ez}. The luminosity
distance $d_L$ is shown by the difference $\Delta(m-M)$ to an empty
($\Omega_{\rm tot}=0$) flat universe. The thin lines show the
accumulated deficit in photons compared to the ``no decoherence'' case
(standard $\Lambda$CDM model) indicated by the thick line.  {\bf Lower
Panel:} As the upper panel, but now showing also a flat CDM model with
$\Omega_m=1$ and $\Omega_\Lambda=0$ (dotted line). For illustration,
we consider a hidden photon model with $p=1$, $N=2$ and
$D_*=6\times10^{-6}$ and show the dimming for the B
($\lambda_\star\simeq440$~nm, upper line) and V
($\lambda_\star\simeq550$~nm, lower line) band. Whereas the overall
dimming effect is practically indistinguishable from the $\Lambda$CDM
model, the strong \emph{reddening} of the starlight from the energy
dependence of $D\propto\omega^2/M_{\rm Pl}$ is challenged by the
data~\cite{KowalskiPC}.}
\label{fig:SNe}
\end{figure}
%%%%%%%%%%%%%%%%%%%%%%%%%%%%%%

\begin{figure*}[ph!]
\includegraphics[width =
0.42\linewidth]{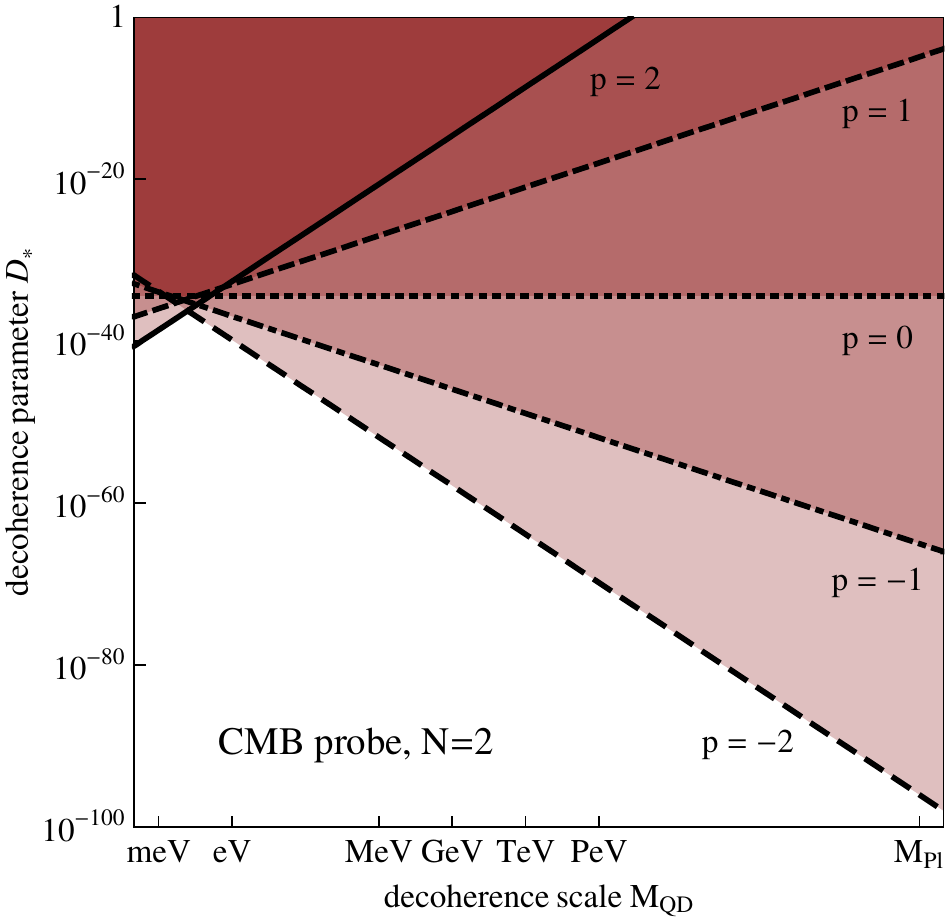}\hspace{1cm}\includegraphics[width =
0.42\linewidth]{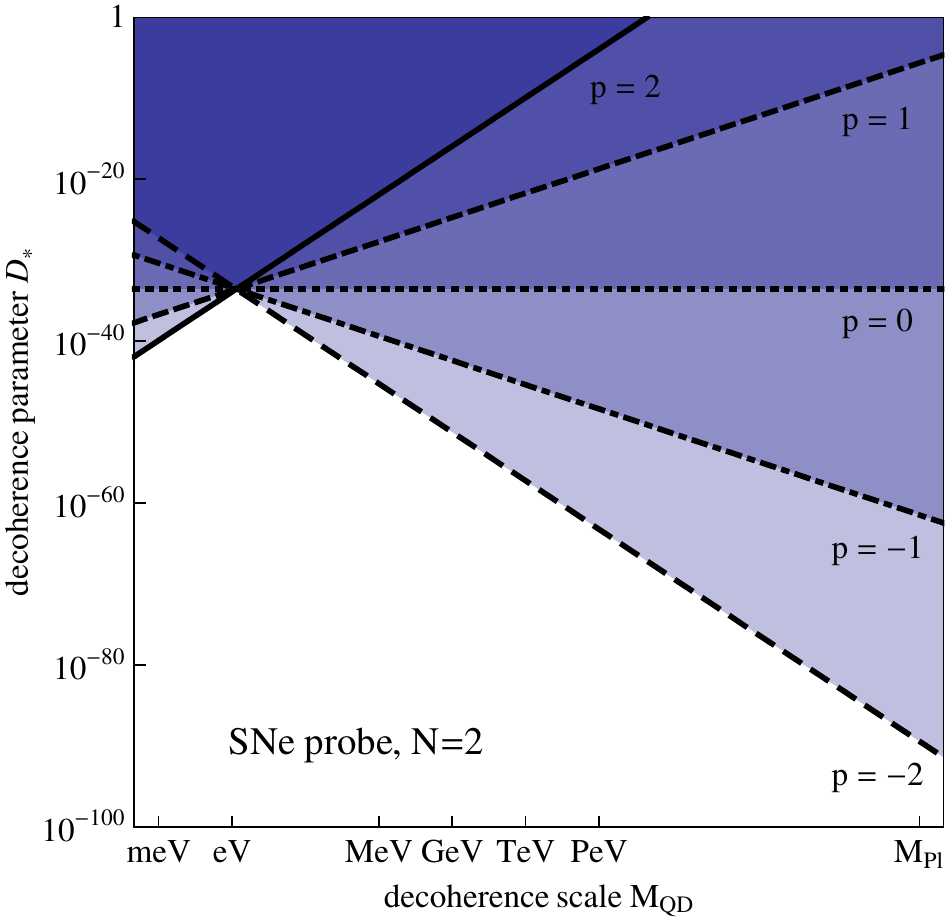}\\[0.2cm]\includegraphics[width
=0.42\linewidth]{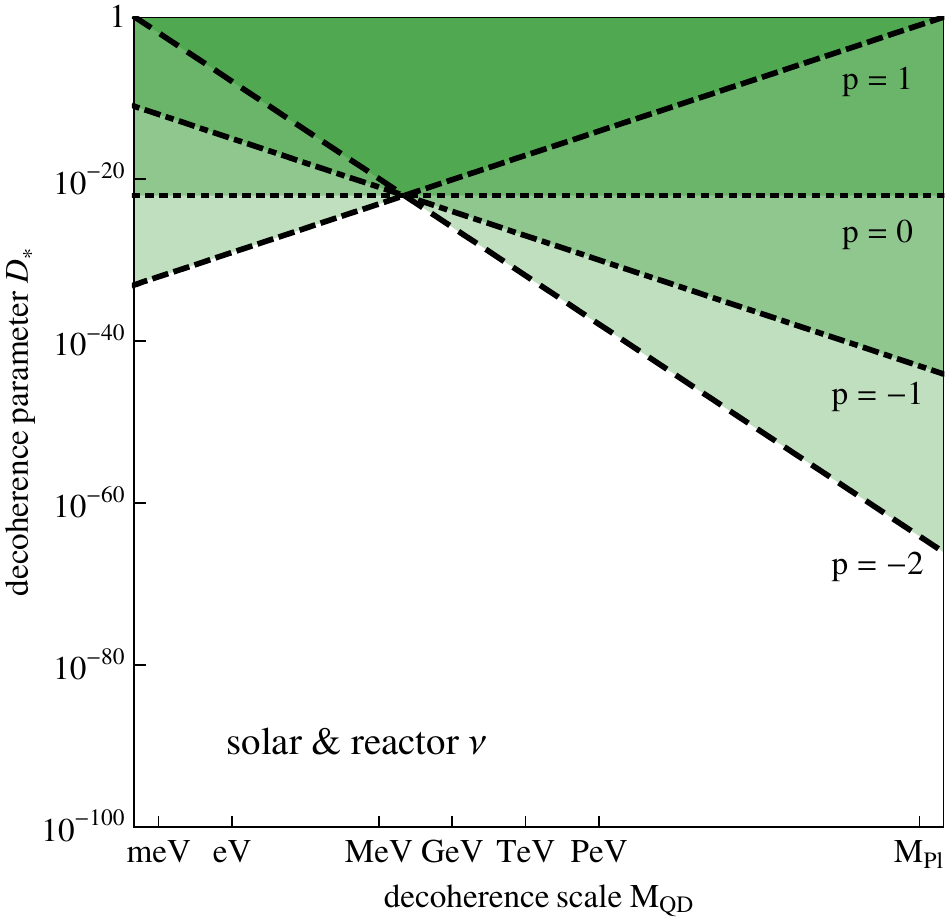}\hspace{1cm}\includegraphics[width
= 0.42\linewidth]{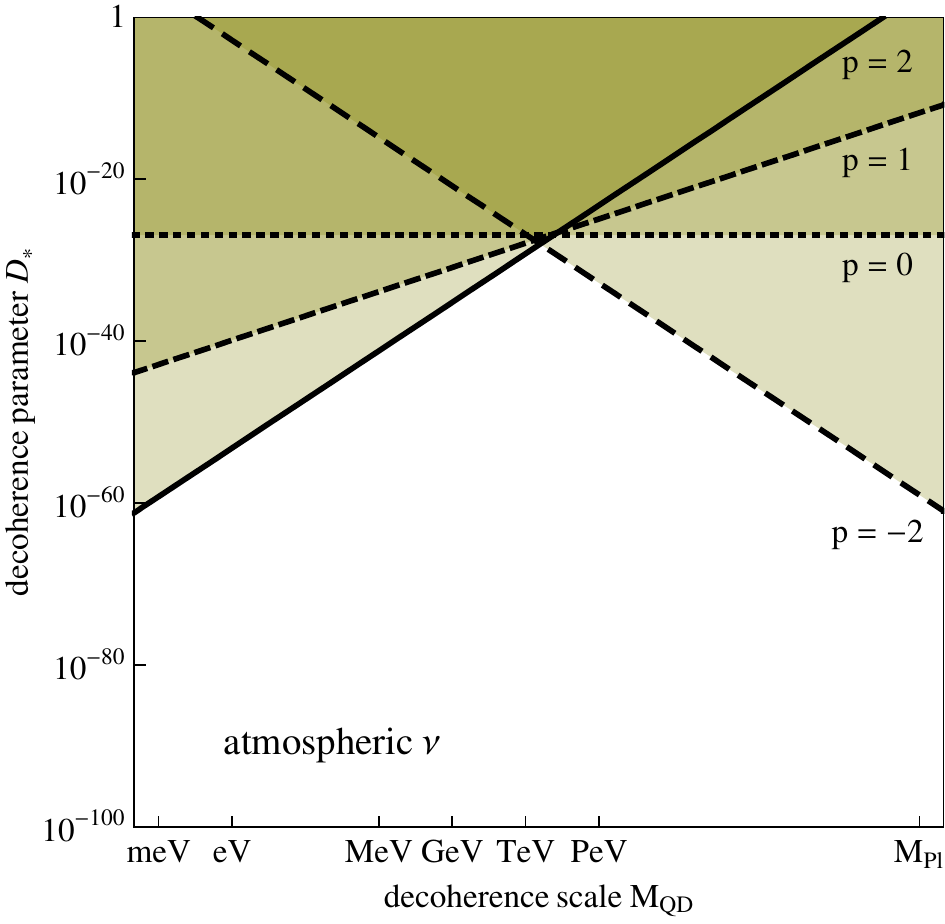}
\caption[]{3$\sigma$ C.L. upper bound on $D_*$ as a function of  
$M_{\rm QD}$ for various models of $D$.
We show the  bounds on from the
CMB~\cite{FIRAS} (upper left panel) and from type Ia
SNe~\cite{Kowalski:2008ez} (upper right panel) 
for various models
$p=0,\pm1,\pm2$ of the decoherence parameter $D=D_*\omega^{p+1}/M_{\rm
QD}^p$. Darker colors indicate regions which are excluded by multiple
models. 
Note that the limits become generally stronger for $N>2$ by up
to a factor $2$. For comparison, the lower plots show the limits from
solar and reactor neutrino data~\cite{Fogli:2007tx} ($3\sigma$ C.L.)
and atmospheric neutrino data,~\cite{Lisi:2000zt} ($3\sigma$~C.L.)
and~\cite{Collaboration:2009nf} ($90$\% C.L.).}
\label{fig:exclusion}
\end{figure*}

As an example, the upper panel of Fig.~\ref{fig:SNe} shows the effect
on quantum decoherence in the $\Lambda$CDM model assuming a single
hidden U$(1)$ ($N=2$).  The luminosity distance of the SNe is shown as
the difference between their measured apparent magnitude $m$ and their
known absolute magnitude $M$,
\begin{equation}
m-M = 5\log_{10}d_{L,{\rm Mpc}}+25\,.
\end{equation} 
As in the previous case we can derive $3\sigma$ limits for various
decoherence models that are shown in Tab.~\ref{tab:threesigma} (see also 
upper right panel in Fig.~\ref{fig:exclusion}). The
limits on Planck-scale suppressed decoherence ($p=1,2$) from SN
dimming are slightly stronger than the CMB limits.

%%%%%%%%%%%%%%%%%%%%%%%%%%%%%%%
%%%%%%%%%%%%%%%%%%%%%%%%%%%%%%%

However, we would like to emphasize that these limits depend on the
cosmological model and the normalization of the SN data. If the SN dimming by photon decoherence is strong
this can have an effect on the evaluation of cosmological data. We
give an example in the lower panel of Fig.~\ref{fig:SNe} where the
decoherence effect could even be able to reproduce the observed SN
luminosities from a CDM model. Note, that this model is not excluded
by the corresponding CMB limits, though it is not compatible with the
analogous bound from atmospheric neutrinos. In addition, one has to
keep in mind that the photon frequency dependence of the parameter $D$
causes a color excess (unless $p=-1$) which is in conflict with
observation~\cite{Kowalski:2008ez}.  For the example shown in the
right panel of Fig.~\ref{fig:SNe} the color excess between the B and V
band of the form, $E[{\rm B-V}] \equiv \Delta(m-M)_{\rm B} -
\Delta(m-M)_{\rm V}$, is larger than $0.8$ for redshifts $0.2\lesssim
z\lesssim0.8$, more than twice the median color excess observed from
SNe at this distance~\cite{KowalskiPC}. Moreover, photon absorption as
a SN dimming mechanism would violate the cosmic distance-duality, {\it
i.e.}~the luminosity and angular diameter distance relation $d_L/d_A =
(1+z)^2$~\cite{Bassett:2003vu}. Hence, it is unlikely that decoherence
can fully account for SN dimming. Nevertheless, it could have an
effect on the evaluation of cosmological data.

The dependence on $M_{\rm QD}$ of the upper bound on $D_*$  
for various models of $D$ as derived  from CMB distortions and SN dimming 
are summarized in the
top panels of Fig.~\ref{fig:exclusion}. The lower panels show
analogous limits derived from solar and reactor neutrino
data~\cite{Fogli:2007tx} as well as atmospheric neutrino
results~\cite{Lisi:2000zt,Collaboration:2009nf}. 
It is apparent from the plots (as the
intersection of the bounds ) that these probes are based on
different energy scales, meV-eV in the case of the CMB, eV in the case
of SNe, multi-MeV in the case of solar and reactor neutrino data and
multi-TeV in the case of atmospheric neutrinos.\footnote{The
sensitivity to astrophysical neutrinos in the TeV-PeV region have also
been discussed in Ref.\cite{Hooper:2004xr}. This probe could
significantly improve the bounds on decoherence for the cases
$p=1,2$.} Consequently we see that 
for models with $p\leq 0$ the bounds from 
from CMB distortions and SN dimming  are the strongest. One must 
however bear in mind that while bounds from neutrino data 
are ``unavoidable'' once the quantum decoherence effects exist,
the effects on photon propagation discussed here rely on the 
extra assumption of the presence of massless hidden $U(1)$'s.  
Besides this, we also note that, generically decoherence effects
that grow with energy are better constrained from higher energy
atmospheric neutrino data. For example for the case $D\propto
\omega^2/M_{\rm QD}$ ($p=1$ in our notation) the possible decoherence
of photons in the presence of a single hidden U$(1)$ is about four
(five) orders of magnitude stronger constrained from CMB (SNe) data
than the corresponding limit from solar and reactor neutrino
data~\cite{Fogli:2007tx} ({\it cf.}~Tab.~\ref{tab:threesigma}) but 
the bound derived from atmospheric
neutrinos~\cite{Collaboration:2009nf} is much stronger.

The dimensionless parameter $D_*$ appearing in our definition of the
decoherence scale~(\ref{eq:GIG}) is naturally expected to be of order
$1$. Hence, atmospheric neutrino data already severely constraints
Planck-scale suppressed decoherence of the form $D=D_*\omega^2/M_{\rm
Pl}$. Similarly, our analogous limits on photon decoherence in the
presence of hidden U$(1)$s require $D_*\ll1$ for $p=1$. In reverse, a
quantum theory of gravity predicting Planck-scale suppressed
decoherence is not compatible with the existence of unbroken hidden
U$(1)$s unless, {\it unnaturally}, $D_*\ll1$.

Higher order quantum decoherence with $p\geq2$ is so far only weakly
constrained by neutrino systems or by photon/hidden photon systems in
the meV to eV energy range. We will speculate in the following
section, how high energy gamma ray sources could possibly explore this
unconstrained parameter space.

\section{Photon Propagation}\label{sec:propagation}

Decoherence effects can also have interesting effects on the
propagation of photons in the presence of photon absorption. We can
account for photon absorption effects in the Liouville equation
(\ref{eq:liouville_mod}) by a contribution
\begin{equation}\label{eq:decay}
\mathcal{D}_{\Gamma}[\rho] =
-\frac{\Gamma}{2}\lbrace\Pi_0,\rho\rbrace\,,
\end{equation}
where $\Gamma = 1/\lambda$ is the photon absorption
rate,\footnote{Since we consider point-source fluxes in the following
we will treat the reaction $\gamma+\gamma_{\rm BG}\to e^++e^-$ as an
absorption process of the photon.  Subsequent electro-magnetic
interactions of secondary $e^\pm$ will contribute to the {\it diffuse}
GeV-TeV background.} {\it e.g.}~in the intergalactic photon background
(BG) via $\gamma+\gamma_{\rm BG}\to e^++e^-$, and $\Pi_0$ is the
photon projection operator. 

The photon survival probability in the presence of a single hidden
U$(1)$ can be readily solved from the modified Liouville equation and
gives (see App.~\ref{app:lindblad} for details)
\begin{multline}\label{eq:Pdecay}
P_{\gamma\to\gamma} =
e^{-\frac{L}{2}(D+\Gamma)}\bigg[\cosh\left(\frac{L}{2}\sqrt{D^2+\Gamma^2}\right)\\-\frac{\Gamma}{\sqrt{D^2+\Gamma^2}}\sinh\left(\frac{L}{2}\sqrt{D^2+\Gamma^2}\right)\bigg]\,.
\end{multline}

We can estimate the sensitivity of extra-galactic TeV gamma ray
sources to decoherence effects as
\begin{equation}
D_* \sim 10^{-38+16p}\left(\frac{M_{\rm QD}}{M_{\rm Pl}}\right)^p
\left(\frac{\rm TeV}{\omega}\right)^{p+1}\left(\frac{\rm
kpc}{L}\right)\,.
\end{equation}
In particular for $M_{\rm QD}\sim M_{\rm Pl}$ , $D_*\sim10^{-22}$ for
$p=1$ or $D_*\sim10^{-6}$ for $p=2$. Hence, this probe has the potential 
to be more sensitive to Planck-scale suppressed decoherence
than existing neutrino data~\cite{Lisi:2000zt,Collaboration:2009nf}
by many orders of magnitude. We will discuss in the following possible
signals of decoherence in the spectra of TeV gamma ray sources.

\begin{figure}[t]
\includegraphics[width = \linewidth]{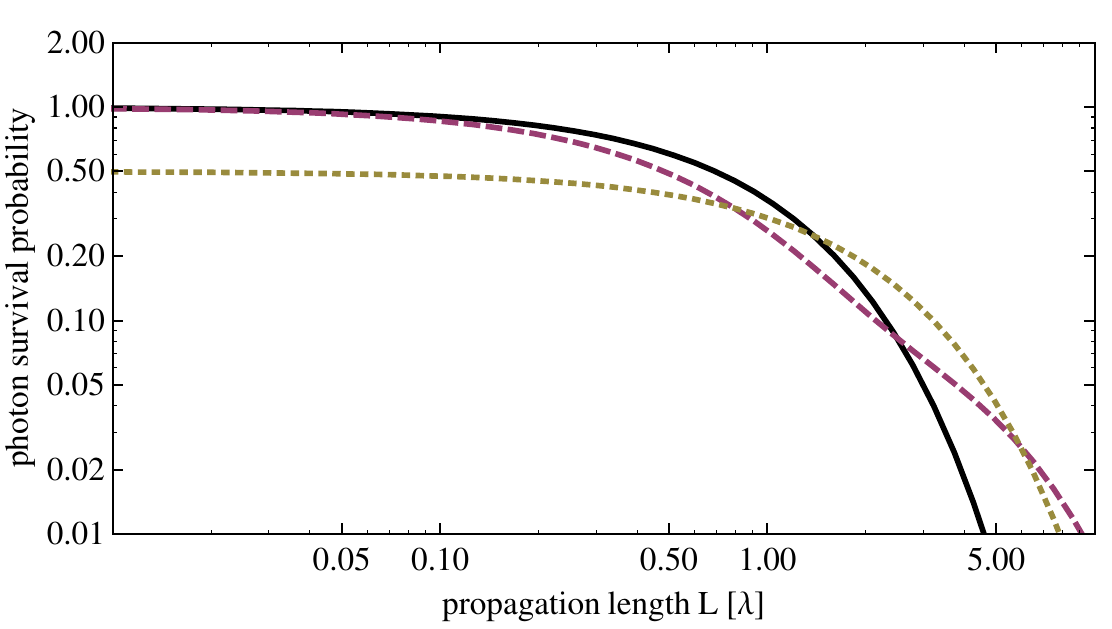}
\vspace{-0.3cm}
\caption[]{Photon survival probability in the presence of
decoherence with a single hidden U$(1)$ ($N=2$). The black line
shows the photon attenuation, {\it e.g.}~in the inter-galactic photon
background via $\gamma+\gamma_{\rm BG}\to e^++e^-$, in the absence of
decoherence ($D=0$). The dashed curve shows decoherence with
$D\lambda=1$ and the dotted curve $D\lambda\gg
1$.}\label{fig:survival}
\end{figure}

Figure~\ref{fig:survival} shows the survival probability for three
different values of $D$. The functional
behaviour can be easily understood as follows: 
The hidden U$(1)$ serves as an invisible ``storage'' of
photons during propagation. If $D\lambda\gg 1$ decoherence quickly equalizes the number of
photons and hidden photons. This results into a \emph{decrease} of the photon 
survival probability to $1/2$ (for $N=2$) for $L\lesssim \lambda$. 
On the other 
hand, inelastic scattering  only affects photons. Therefore 
at larger $L$ photons can be ``replenished'' by the decoherence
of the unabsorbed hidden photons into photons. Hence, for propagation distances
$L>\lambda$ the presence of hidden photons \emph{increases} the photon
survival probability.

For the general case of $N$ U$(1)$s and strong decoherence
$D\gg\Gamma$ we can express the photon survival probability at a
distance much larger than the decoherence scale $L\gg1/D$ as (see
App.~\ref{app:lindblad} for details)
\begin{equation}\label{eq:Pasym}
P^{}_{\gamma\to\gamma} \simeq
\frac{1}{N}\exp\left(-\frac{L}{N\lambda}\right)\,.
\end{equation}
This can have an important effect on the spectra of TeV gamma ray
sources.\footnote{Note that in the absence of hidden sectors, as long as
$\omega \lesssim v$ ($v$ being the scale of electroweak symmetry
breaking )  the photon survival probability is 1 independently on
whether the Higgs potential around virtual black holes has its minimum
at the trivial vacuum $v = 246$~GeV or at the unbroken vacuum $v =
0$. For photon energies beyond the electroweak breaking scale,
one may theorize over  possible decoherence effect between the
neutral SM gauge bosons. In this case 
 the conservation of energy and momentum in decoherence effects, which
forbids transitions of the form $\gamma\to Z$, is a crucial
assumption.}  Firstly, if the onset of decoherence appears at energies
covered by the spectra one could observe a step-like drop of the flux
by a factor $1/N$. And secondly, as long as the absolute source
emissivity of photons is unknown (such that the pre-factor $1/N$ in
Eq.~(\ref{eq:Pasym}) gets renormalized), the expected spectral
cut-offs of the sources could be shifted according to an extend photon
interaction length $\lambda_{\rm eff}=N\lambda$. These effects can be
clearly seen in Fig.~\ref{fig:survival} for the case $N=2$. For a TeV
gamma ray source at $100$~Mpc and Planck-scale suppressed decoherence
this requires $D_*\gg
10^{-27}$ ($D_*\gg 10^{-11}$) for $p=1$ ($p=2$). In comparing this
with Tab.~\ref{tab:threesigma} and Fig.~\ref{fig:exclusion} we see
that there is ample room for models that could have such an effect on
the spectra.

%%%%%%%%%%%%%%%%%%%%%%%%%%%%%%%
%%%%%%%%%%%%%%%%%%%%%%%%%%%%%%%

\section{Conclusions}\label{sec:summary}

We have discussed the effects of quantum decoherence of photons in the
presence of hidden sector U$(1)$s. Quantum decoherence in the system
of photon and hidden photons could be induced by a foam-like structure
of space-time in a quantum theory of gravity, where virtual black
holes pop in and out of existence at scales allowed by Heisenberg's
uncertainty principle.

We have shown that these decoherence effects are strongly constrained
by the absence of photon disappearance in the cosmic microwave
background.  Furthermore quantum decoherence as an additional source
of starlight dimming can also be constrained by the luminosity
distance measurements of type Ia supernovae. Consequently based on the
standard $\Lambda$CDM model we can derive constraints on the
decoherence in the presence of hidden U$(1)$s.  In principle, the
decoherence effect could be strong enough to influence the evaluation
of cosmological data. However, color dependencies in the dimming via
photon decoherence are not favored by the data and could be used to
derive further constraints.

Our main results are summarized in  the upper panel of
Fig.~\ref{fig:exclusion} and compared with the present bounds
from non observation of decoherence effects in neutrino systems. 
For models with decoherence parameter $D$ constant or decreasing
with energy,  the bounds from from CMB distortions and SN dimming  
are the strongest. One must 
however bear in mind that while bounds from neutrino data 
are ``unavoidable'' once the quantum decoherence effects exist,
the effects on photon propagation discussed here rely on the 
extra assumption of the presence of massless hidden U$(1)$s.  
We also note that, generically decoherence effects
which grow with energy are better constrained from the higher energy
atmospheric neutrino data.

We have also discussed the interplay between photon absorption and
decoherence.  This effect can become important for distant TeV gamma
ray point-sources if the decoherence length is much smaller than the
photon interaction length. Assuming $N-1$ additional U$(1)$s this
could leave characteristic features in gamma ray spectra in the form
of step-like drops by factors $1/N$ or by an effective increase of the
absorption length $\lambda$ to $\lambda_{\rm eff} = N\lambda$.

\section*{Acknowledgments}
We would like to thank Francis Halzen and Andreas Ringwald for comments on the manuscript and Marek Kowalski for his help on the discussion of model constraints from a limited color excess in SN data.
This work is supported by 
US National Science Foundation Grant No  
PHY-0757598 and PHY-0653342,   
by The Research Foundation of SUNY at Stony
and the UWM Research Growth Initiative. M.C.G-G acknowledges further
support from  Spanish MICCIN grants 2007-66665-C02-01 and 
consolider-ingenio 2010 grant CSD2008-0037 and 
by CUR Generalitat de Catalunya grant 2009SGR502

\appendix

\section{Lindblad Formalism}\label{app:lindblad}

We outline the solution to the Liouville Eq.~(\ref{eq:liouville_mod})
in the presence of quantum decoherence. The density matrix $\rho$ and
Lindblad operators $b_j$ can be expanded in a basis of hermitian
matrices $F_\mu$ that satisfy the orthonormality condition ${\rm
Tr}(F^\dagger_\mu F_\nu) = \delta_{\mu\nu}/2$. Without loss of
generality we consider a basis with
$(F_0)_{ij}=\delta_{ij}/\sqrt{2N}$. Explicitly, we have
\begin{equation}
\rho=\sum_\mu \rho_\mu F_\mu\,,\qquad b_j=\sum_\mu b^{(j)}_\mu
F_\mu\,.
\end{equation}
The free propagation of photons and hidden photons ($H = -{\rm
i}\partial/\partial x$) can be readily solved in terms of
``light-cone'' coordinates $\hat{x} = (x-t)/2$ and $\hat{t} =
(x+t)/2$. In these new coordinates Eq.~(\ref{eq:liouville_mod}) can be
written $\partial\rho/\partial\hat{t} = \mathcal{D}[\rho]$.

Hence, the coefficients of the free equations of motion satisfy the
differential equation
\begin{equation}
\frac{\partial}{\partial t}\rho_\mu = - \sum_\nu D_{\mu\nu}\rho_\nu\,,
\end{equation}
with $D_{\mu0}=D_{0\mu}=0$ and
\begin{equation}\label{defDij}
D_{ij} = \frac{1}{2}\sum_{k,l,m,n}b^{(n)}_mf_{iml}b^{(n)}_kf_{jkl}\,,
\end{equation}
where $f_{ijk}$ are structure constants defined by $[F_i,F_j] =
i\sum_kf_{ijk}F_k$.

The solution of $\partial_t\rho_0=0$ is trivial and requires
$\rho_0(t)=\rho_0=2{\rm Tr}(\Pi_\alpha F_0)=\sqrt{2/N}$ for all
species $\alpha$. If $D_{ij}$ is diagonalizable by a matrix $M$,
$(M^{-1}DM)_{ij} = D_{i}\delta_{ij}$, we can write the final solution
as
\begin{equation}
P_{\gamma\to\gamma} = \frac{1}{N}+\frac{1}{2}\sum_{i,k,j}e^{-D_k
t}\rho_i(0)M_{ik}M^{-1}_{kj}\rho_j(0)\,.
\end{equation}
This reduces to Eq.~(\ref{eq:Pgg}) in the case $D_i=D$ using
$2=\sum_\mu\rho_\mu^2$ following from $\rho^2(0)=\rho(0)$ and ${\rm
Tr}[\rho]=1$.

We can extend the Lioville Eq.~(\ref{eq:liouville_mod}) by a photon
decay term of the form~(\ref{eq:decay}). The modified equation can be
readily solved in the case of $N=2$, taking $F_i =
\frac{1}{2}\sigma_i$ with Pauli matrices $\sigma_i$. In this basis the
photon projection operator in the term~(\ref{eq:decay}) has the form
$\Pi_0 = F_0+F_3$. And the general solution of the photon survival
probability is given in Eq.~(\ref{eq:Pdecay}).

In the case of strong decoherence, {\it i.e.}~at propagation distances
$L\gg1/D$ and $D\gg\Gamma$, we can derive the asymptotic
solution~(\ref{eq:Pasym}) of the general survival probability with $N$
U$(1)$s in the following way. In the presence of strong decoherence we
can assume that at any step during the evolution ${\rm
Tr}[\Pi_\mu\rho] \simeq {\rm Tr}[\Pi_\nu\rho]$ and, in particular,
${\rm Tr}[\rho] \simeq N{\rm Tr}[\Pi_0\rho]$. Since the trace of the
decoherence term vanishes we can derive the asymptotic differential
equation
\begin{equation}
{\rm Tr}[\dot\rho] \simeq N{\rm Tr}[\Pi_0\dot\rho] \simeq -\Gamma{\rm
Tr}[\Pi_0\rho]\,.
\end{equation}
This has the solution (\ref{eq:Pasym}), noting that
$P_{\gamma\to\gamma}={\rm Tr}[\Pi_0\rho]$ and initially ${\rm
Tr}[\Pi_0\rho(0)] = 1/N$.

\end{document}